\documentclass[12pt]{article}
\pdfoutput=1
\usepackage{setspace,braket}
\usepackage{amsmath, amssymb, amsthm, float,graphicx}
\def\K{{\mathcal K}}

\def\be{\begin{equation}}
\def\ee{\end{equation}}
\def\lp{\ell_P}
\def\R{{\mathcal R}}
\def\R{{\mathcal R}}

\def\K{{\mathcal K}}
\def\be{\begin{equation}}
\def\ee{\end{equation}}
\def\lp{\ell_P}

\def\a{\alpha}
\def\b{\beta}
\def\g{\gamma}

\def\r{\rho}

\def\m{\mu}\def\n{\nu}\def\s{\sigma}

\def\beq{\begin{eqnarray}}\def\eeq{\end{eqnarray}}
\def\tr{{\rm tr~}}

\begin {document}

\title{\bf  Constraining gravity using entanglement in AdS/CFT }
\date{}

\author{Shamik Banerjee\footnote{On leave of absence from Kavli IPMU.}, Arpan Bhattacharyya, Apratim Kaviraj, Kallol Sen \\ \& Aninda Sinha\\ ~~~~\\
\it Centre for High Energy Physics,
\it Indian Institute of Science,\\ \it C.V. Raman Avenue, Bangalore 560012, India. \\}
\maketitle

\vskip 1cm
\begin{abstract}{\small We investigate constraints imposed  by entanglement on gravity in the context of holography. 
First, by demanding that relative entropy is positive and using the Ryu-Takayanagi entropy functional, we find certain constraints at a nonlinear level for the dual gravity.  Second, by considering Gauss-Bonnet gravity, we show that for a class of small perturbations around the vacuum state, the positivity of the two point function of the field theory stress tensor guarantees the positivity of the relative entropy. Further, if we impose that the entangling surface closes off smoothly in the bulk interior, we find restrictions on the coupling constant in Gauss-Bonnet gravity.  We also give an example of an anisotropic excited state in an unstable phase with broken conformal invariance which leads to a negative relative entropy. }
\end{abstract}
\tableofcontents
\onehalfspace
\section{Introduction}

In recent times, there has been a huge interest to see what quantum entanglement \cite{Bombelli:1986rw,  ryu, calabrese, rev, mukund, chm, maldacena, myersme, Hubeny:2012wa, wall} can teach us about gravity. Certain entanglement measures such as relative entropy \cite{nc}, which roughly speaking tells us how distinguishable two states are, need to be positive in a unitary theory. The positivity of this quantity was studied in holographic field theories with two derivative gravity duals in \cite{Relative}--related work include \cite{Nozaki:2013vta}. In the context of quantum field theories with holographic dual gravity descriptions, one can ask what this inequality translates into. Furthermore, the Ryu-Takayanagi prescription \cite{ryu} (and its extensions to more general gravity theories \cite{higher, higher2, ab, solo, abs, dong} ) gives us a way to compute how the entangling surface extends into the bulk. For the vacuum state in a conformal field theory, one may expect that the surface closes off smoothly in the interior as can be checked by explicit calculations in Einstein gravity. By demanding a smooth surface in the bulk, we can try to see if a general theory of gravity gets constrained. These will be the main questions of interest in this paper.

Let us begin by discussing relative entropy. Relative entropy between two states $\rho$ and $\sigma$ is defined as
\be
S(\rho|\sigma)= \tr (\rho \log \rho)-\tr (\rho \log \sigma)\,.
\ee
As reviewed in appendix A, in quantum mechanics, this quantity is positive for a unitary theory. In \cite{Relative}, relative entropy was discussed in the  holographic context. The state $\sigma$ was chosen to be the reduced density matrix for a spherical entangling surface. In this case, $\sigma\equiv e^{-H}/\tr e^{-H}$ with $H$ being the modular hamiltonian. It can be easily shown that (see eg.\cite{cas1,Relative})
\be\label{dhds}
S(\rho|\sigma)=\Delta H-\Delta S\,,
\ee
where $\Delta H=\langle H\rangle_1 -\langle H\rangle_0$ and $\Delta S= S(\rho)-S(\sigma)$ with $S(\rho)=-\tr \rho \log \rho$ being the von Neumann entropy for $\rho$ and is the entanglement entropy for a reduced density matrix $\rho$. Then the positivity of $S(\rho|\sigma)$ would require,
\be
\Delta H\ge \Delta S\,.
\ee
Now we can calculate the modular hamiltonian for the sphere \cite{Haag}, from the formula,
\be\label{modH}
H=2\pi \int_{r<R}d^{d-1} x \frac{R^2-r^2}{2R}T_{00}\,.
\ee
Here $T_{\mu\nu}$ is the $d$-dimensional field theory stress tensor and $00$ is the time-time component. We know how to compute $T_{\mu\nu}$ in holography. The Ryu-Takayanagi prescription (and known generalizations) gives us a way to compute $\Delta S$. Thus we can check if  and how the inequality $\Delta H\geq \Delta S$ is satisfied. In \cite{Relative}, many examples were considered and in each case it was shown that this inequality is respected in Einstein gravity. 
If we consider a small excitation around the vacuum state then to linear order in the perturbation $\Delta H=\Delta S$. This can be shown to be equivalent with the linearized Einstein equations \cite{einstein1, einstein2}. This equality has been recently shown to hold for a general higher derivative theory of gravity \cite{einstein3}. It is thus very interesting to ask what constraints we get at the nonlinear order. We will address this question for the special case of a constant stress tensor for the case where the holographic entanglement entropy is given by the Ryu-Takayanagi prescription---in other words, we will ask if even at non-linear level we get Einstein gravity. We find that the constraints arising from relative entropy give us a larger class of models than just Einstein gravity. However, we show that there exists matter stress tensor for which the bulk null energy condition is violated everywhere except at the Einstein point. This in turn implies that relative entropy can continue to be positive although the bulk null energy condition is violated. In fact we can ask the question the other way round: are there examples where the relative entropy is negative but the bulk null energy condition still holds? We will give an example where this happens. Thus the connection between energy conditions and the positivity of relative entropy, which in some sense is reminiscent of the connection between energy conditions and the laws of thermodynamics, appears to be less direct than what one would have expected.

In order to get some intuition about what feature of gravity ensures the positivity of relative entropy, we extend the calculations in \cite{Relative} to higher derivative theories. In particular we focus on Gauss-Bonnet gravity in 5 bulk dimensions \cite{gbholo1,gbholo,Camanho:2013pda} since in this context there is a derivation  \cite{ab,abs} of the corresponding entropy functional \cite{jm, higher, higher2}. We find that in all examples that we consider, the positivity of the two point function of the stress tensor guarantees that the relative entropy is positive. In particular we show this for a constant field theory stress tensor as well as for a disturbance that is far from the entangling surface. 

The inequality for relative entropy can only be explicitly checked when the modular hamiltonian is known. Unfortunately, currently this is not known for cases when the entangling region is a cylinder or a slab. 
From explicit calculations in the context of Einstein gravity, it is known that the entangling surfaces for sphere, cylinder and the slab in the bulk corresponding to the vacuum state close off smoothly. The entangling surface equation gets modified in the presence of higher derivative corrections. The smoothness of the surface imposes constraints on the higher derivative coupling constants. For the slab, this question was addressed in \cite{Ogawa:2011fw}. We will extend this calculation to the other two cases and find bounds on the Gauss-Bonnet coupling constant. We will find that this simple criterion leads to somewhat weaker constraints on the coupling as compared to what arises from micro-causality considerations \cite{micro1, hofman}. However, quite curiously, the bounds are in good agreement with the $a/c$ bounds \cite{hofman} for a non-supersymmetric field theory.

This paper is organized as follows. In section 2, we consider constraints arising from the positivity of relative entropy in a holographic set up where the entanglement entropy is given by the Ryu-Takayanagi entropy functional. These constraints arise at a quadratic order in a perturbation with a constant field theory stress tensor. In section 3, we turn to the study of relative entropy in Gauss-Bonnet holography. In section 4, we investigate the relative entropy for an anisotropic plasma which breaks conformal invariance. We find that the relative entropy in this case is negative and we suggest some possible explanations for this. In section 5, we derive constraints on the Gauss-Bonnet coupling by demanding that the entangling surface extending into the bulk closes off smoothly. We conclude with open problems in section 6. The appendices contain further calculations relevant for the rest of the paper. We will use capital latin letters to indicate bulk indices and greek letters to indicate boundary indices.  Lower case latin letters will indicate an index pertaining to the co-dimension 2 entangling surface.\\
\noindent{\bf Note added:} The paper by Erdmenger et al \cite{erd} which appeared on the same day in the arXiv, deals with a related idea of looking for pathological surfaces in certain higher derivative theories of gravity.

\section{Relative entropy considerations}
In this section we will use the results in \cite{Relative} to derive certain constraints at nonlinear order that arise due to the positivity of relative entropy. In Fefferman-Graham coordinates, the bulk metric can be written as 
\be\label{fgmetric}
ds^2=\frac{L^2}{z^2} dz^2+g_{\mu\nu}dx^\mu dx^\nu\,.
\ee
For Einstein gravity, the bulk equations of motion allow us to systematically solve for $g_{\mu\nu}$ as an expansion around the boundary $z=0$ (see eg.\cite{skenderis}). The idea here is to see what mileage we get if we do not know what the bulk theory is but we demand that the relative entropy calculated using the Ryu-Takayanagi entropy functional is positive.
We want to calculate the quadratic correction to the entanglement entropy for the following form of boundary metric,
\be\label{metricpert}
g_{\mu\nu}=\frac{L^2}{z^2}\left[\eta_{\mu\nu}+a z^d T_{\mu\nu} +a^2 z^{2d}(n_1T_{\mu\alpha}T_\nu^\alpha+n_2\ \eta_{\mu\nu}T_{\alpha\beta}T^{\alpha\beta})+\cdots\right]\,,
\ee
where $a=\frac{2}{d}\frac{\lp^{d-1}}{\hat{L}^{d-1}}$. This form is consistent with  Lorentz invariance  for a constant $T_{\mu\nu}$. We will treat $T_{\mu\nu}$ as a small perturbation to the vacuum. At linearized order, it has been shown in \cite{einstein1,einstein2,einstein3} Einstein equations arise from the condition $\Delta H=\Delta S$. We wish to investigate what happens at the next order. We will keep $n_1$ and $n_2$ arbitrary and derive constraints on them arising from the inequality $\Delta H \geq \Delta S$. Our analysis follows \cite{Relative} very closely, the only change being that we will not specify $n_1$ and $n_2$ to be the Einstein values. Since at linear order (the argument will be reviewed in the next section) we have the equality $\Delta H=\Delta S$ and since $T_{00}$ from the holographic calculation is just given by the coefficient of the $z^d$ term in the metric, the inequality implies $\Delta S\leq 0$ at quadratic order. Thus our task is to calculate $\Delta^{(2)}S$, the quadratic correction to $\Delta S$, as a function of $n_1, n_2$. The analysis below is valid for $d>2$.


\noindent We start with the Ryu-Takayanagi prescription for calculating entanglement entropy in holography,
\be\label{RT}
S=\frac{2\pi}{\lp^{d-1}}\int d^{d-1}x\sqrt{h}\,.
\ee
From Taylor expansion one can show that the quadratic correction to $\sqrt{h}$ is,
\be
\delta^{(2)}\sqrt{h}=\frac{1}{8}\sqrt{h}(h^{ij}\delta h_{ij})^2+\frac{1}{4}\sqrt{h}\ \delta h^{ij} \delta h_{ij}+\frac{1}{4}\sqrt{h}\ h^{ij}\delta^{(2)} h_{ij}\,.
\ee
The induced metric is,
\be
h_{ij}=g_{ij}+\frac{L^2}{z^2}\partial_i z \partial_j z \,.
\ee
This is evaluated at the extremal surface $z=z_0+\epsilon z_1=\sqrt{R^2-r^2}+\epsilon z_1$. Hence, at 0-th order, the metric and its inverse are,
\be
h_{ij}=\frac{L^2}{z_0^2}\left( \eta_{ij}+ \frac{x_ix_j}{z_0^2} \right)\hspace{1cm}\mbox{and}\hspace{1cm}h^{ij}=\frac{z_0^2}{L^2}\left( \eta^{ij}- \frac{x^ix^j}{R^2}\right)\,.
\ee
In $\Delta^{(2)}S$, we get 3 kinds of second order contributions.  To be systematic, we write,
\be\label{3cont}
\int d^{d-1}x\ \delta^{(2)}\sqrt{h}=A_{(2,0)}+A_{(2,1)}+A_{(2,2)}\,,
\ee
where schematically, these are the $(\delta g)^2$, $z_1\delta g $ and $z_1^2$ contributions respectively. To calculate the first term, we can set $z_1=0$. Then 
\be
\delta h_{ij}=aL^2 z^{d-2} T_{ij}  \hspace{1cm}\mbox{and}\hspace{1cm}\frac{\delta^{(2)}h_{ij}}{2}=a^2L^2z_0^{2d-2}(n_1T_{i\alpha}T_j^\alpha+n_2\ \eta_{ij}T_{\alpha\beta}T^{\alpha\beta}) \,.
\ee
This gives,
\begin{multline}
A_{(2,0)}=L^{d-1}a^2\int d^{d-1}x\ Rz_0^d \left( T_{i0}T^{i0}\left( \frac{n_1}{2}+(d-1)n_2-n_2 \frac{r^2}{R^2} \right)+{(T_{00})}^2\left( \frac{n_2}{2}(d-1)-\frac{n_2r^2}{2R^2} \right)\right.\\\left.+T_{ij}T^{ij}\left( \frac{n_1}{2}+\frac{n_2}{2}(d-1)-\frac{n_2r^2}{2R^2} -\frac{1}{4}\right)-\frac{n_1}{2R^2}x^ix^j T_{i0}T^0_j+x^ix^j T_{ik}T^k_j\left(\frac{1}{2R^2}- \frac{n_1}{2R^2} \right)+\frac{1}{8}\left( T^2-T_x^2-2TT_x \right)\right)\,, \nonumber\\ 
\end{multline}
where $T_x=x^ix^j\frac{T_{ij}}{R^2}$ and $T=T_i^i$.
The last two terms in (\ref{3cont}) are same as they appear in \cite{Relative} . Quoting the result,
\be
A_{(2,1)}=L^{d-1}a\int d^{d-1}x \frac{R}{2z_0}\left[ T\left( z_1-\frac{z_0^2}{R^2}x^i\partial_i z_1 \right)+\frac{T_{ij}}{R^2}\left( 2z_0^2x^i\partial^jz_1-z_1x^ix^j-\frac{z_0^2x^ix^jx^k\partial_kz_1}{R^2} \right) \right]\,,
\ee
\be
A_{(2,2)}=L^{d-1}\int d^{d-1}x\frac{R}{z_0^d}\left[ \frac{d(d-1)z_1^2}{2z_0^2}+\frac{z_0^2(\partial z_1)^2}{2R^2}-\frac{z_0^2 (x^i\partial_iz_1)^2}{2R^4}+\frac{(d-1)x^i\partial_iz_1^2}{2R^2} \right]\,.
\ee
We can find $z_1$ by minimizing $A_{(2,1)}+A_{(2,2)}$, which gives,
\be
z_1=-\frac{aR^2z_0^{d-1}}{2(d+1)}(T+T_x)\,.
\ee
Plugging this and summing we get from eq.(\ref{3cont}),
\be\label{7coeff}
\int d^{d-1}x\ \delta^{(2)}\sqrt{h}=L^{d-1}a^2\int d^{d-1}x\ \left( c_1T^2+c_2T_x^2+c_3T_{ij}^2+c_4T_{i0}T^{i0}+c_5\frac{x^ix^jT_{ik}T_j^k}{R^2}+c_6\frac{x^ix^jT_{i0}T_j^0}{R^2}+ c_7TT_x \right)\,,
\ee
where unlike \cite{Relative}\footnote{There appears to be an overall sign missing for $c_6$ in  \cite{Relative}.}, the coefficients $c_1\cdots c_7$ are dependent on $n_1$ and $n_2$,

\begin{eqnarray}
c_1&=&\frac{(R^2-r^2)^{(d-4)/2}}{8(1+d)^2R}\left(-4(1+d)^2n_2(r^2-R^2)^2(r^2-(d-1)R^2)\right.\\ &&\ \ \ \ \left.+R^2(2(d^2+2d-1)r^4+(1-5d^2)r^2R^2+(2d^2-d-1)R^4) \right)\,,\\
c_2&=&\frac{\left(-r^2+R^2\right)^{\frac{1}{2} (-4+d)} \left(\left(1-5 d^2\right) r^2 R^3+(-3+d (3+4 d)) R^5\right)}{8 (1+d)^2}\,,\\
c_3&=&\frac{\left(-r^2+R^2\right)^{d/2} \left(-2 n_2 r^2+(-1+2 n_1+2 (-1+d) n_2) R^2\right)}{4 R}\,,\\
c_4&=&\frac{\left(-r^2+R^2\right)^{d/2} \left(n_1 R^2-2 n_2 \left(r^2-(-1+d) R^2\right)\right)}{2 R}\,,\\
c_5&=&\frac{\left(d^2-(1+d)^2 n_1\right) R \left(-r^2+R^2\right)^{d/2}}{2 (1+d)^2}\,,\\
c_6&=&-\frac{n_1}{2}  R \left(-r^2+R^2\right)^{d/2}\,,\\
c_7&=&\frac{(-1+d) R^3 \left(-r^2+R^2\right)^{\frac{1}{2} (-4+d)} \left((1-3 d) r^2+(1+2 d) R^2\right)}{4 (1+d)^2}\,.
\end{eqnarray}
Now we integrate the expression (\ref{7coeff}) over the $(d-2)$-sphere on the boundary. We use the trick,
\be
\int d^{d-1}x \ f(r)x^ix^jx^kx^l\cdots \mbox{$n$ pairs}=N(\delta_{ij}\delta_{kl}\cdots+\mbox{permutations})\int d^{d-1}x \ f(r) r^{2n}\,,
\ee
where $N$ is some normalization constant. For $n=1$, $N=1/(d-1)$; and for $n=2$, $N=1/((d-1)^2+2(d-1))$. The final result comes out in the form \footnote{The expression for $C_3$ in \cite{Relative} after substituting for $n_1, n_2$ is off by a factor of $d/(d+2)$ although the overall sign is correct. This appears to be related to the opposite sign used for $c_6$. We have cross-checked our results on mathematica for various cases and the notebook may be made available on request.},
\be
\int d^{d-1}x\sqrt{h}=a^2L^{d-1}\Omega_{d-2}\left(
C_1T^2+C_2T_{ij}^2+C_3T_{i0}^2 \right)\,,
\ee
with
\be
C_1=\frac{2^{-3-d}d\left(1+4\left(d^2-1\right)n_2\right)\sqrt{\pi }R^{2d}\Gamma [d+1]}{\left(d^2-1\right)\Gamma \left[\frac{3}{2}+d\right]}\,,
\ee
\be
C_2=\frac{2^{-3-d} d  \sqrt{\pi } R^{2 d} \Gamma[1+d]}{\left(d^2-1\right) \Gamma\left[\frac{3}{2}+d\right]}\left(-1-2 d+4 (d+1) n_1 +4 \left(d^2-1\right) n_2\right)\,,
\ee
\be
C_3=-\frac{2^{-1-d} d (n_1+2 (d-1) n_2) \sqrt{\pi } R^{2 d} \Gamma[1+d]}{(d-1) \Gamma\left[\frac{3}{2}+d\right]}\, .
\ee

Now we must demand that $\Delta^{(2)}S\leq 0$. We can write $\Delta^{(2)}S= V^T M V$ with $V$ being a $(d-1)(d+2)/2$ dimensional vector with the independent components of $T_{\mu\nu}$ as its components. Then demanding that the eigenvalues of $M$ are $\leq 0$ will ensure $\Delta^{(2)} S\leq 0$. This leads to 
\begin{eqnarray}
n_1+2(d-1)n_2 &\geq& 0 \,,\\
2d+1-4(d+1)n_1-4(d^2-1)n_2 &\geq& 0\,,\\
d+2-4(d+1)n_1-4d(d^2-1)n_2 &\geq& 0\,.
\end{eqnarray}
We get the region indicated in fig.1 allowed by this set of inequalities. One interesting observation is that when $d\rightarrow \infty$, then the allowed region becomes the interval $0\leq n_1\leq 1$ with $n_2=0$ coinciding with the Einstein result. The area of the triangle is given by 
\be
{\rm Area_d}= \frac{d^2}{8 (d+1)^2(d-2)}\,.
\ee
Notice that the (extrapolated) Area$_{d=2}$ is infinity. This makes sense since in $d=2$ we expect constraints on only 2 eigenvalues (since $T^2$ and $T_{ij}^2$ are no longer independent) which will give us an unbounded region. Further Area$_{d\rightarrow \infty} \rightarrow 0$ which leads to a line interval for $d\rightarrow \infty$ as shown in fig.1.

At this stage, we have a wider class of theories that are allowed by the inequality than the Einstein theory. The other theories need extra matter in addition to Einstein gravity to support them. As such we could ask if the matter needed satisfies the null energy condition. 

\begin{figure}[ht]
\centering
\includegraphics[scale=.6]{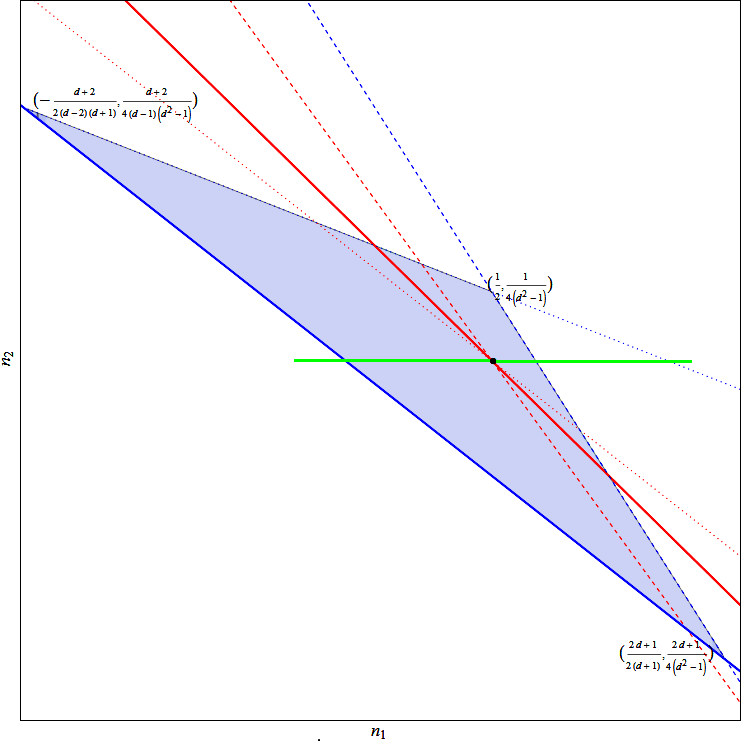} 
\caption{(colour online) For $d>2$ we get the allowed $n_1,n_2$ region to be the blue triangle above for a generic stress tensor. The region above the blue solid line and below the blue dashed and dotted lines are allowed from the relative entropy positivity. For $d\rightarrow \infty$ the region collapses to a line $0\leq n_1\leq 1$ indicated in green. The Einstein value $(n_1,n_2)=(\frac{1}{2},-\frac{1}{8(d-1)})$ is shown by the black dot. The region below the solid red line and above the dashed and dotted red lines are allowed by the null energy condition. By turning on a generic component of the stress tensor only the Einstein value is picked out. By switching off certain components of the stress tensor, various bands bounded by the solid, dashed and dotted lines are picked out.}
\end{figure}

As an example consider turning on a constant $T_{01}$ in $d=4$. Then we find
\be
R_{AB}-\frac{1}{2}g_{AB}(R+\frac{12}{L^2})=T^{bulk}_{AB}\,,
\ee
with $T^{bulk}_{AB}$ working to be
\be
T^{bulk}_{AB}=16 z^6 T_{01}^2 \left[\frac{3}{2} (\delta n_1+4\delta n_2)\delta_A^z\delta_B^z +  (\delta n_1+6\delta n_2) \delta_A^0\delta_B^0- (\delta n_1+6\delta n_2) \delta_A^1\delta_B^1-2(\delta n_1+3\delta n_2) \sum_{i=2,3}\delta_A^i\delta_B^i\right]\,.
\ee
Here $\delta n_1=n_1-1/2$ and $\delta n_2=n_2+1/24$. 
Using this we find that the null energy condition $T^{bulk}_{AB}\zeta^A \zeta^B \geq 0$ leads to 
\begin{eqnarray}
T^{bulk}_{00}+T^{bulk}_{22}&=&-\delta n_1 \geq 0\,,\\
T^{bulk}_{00}+T^{bulk}_{zz}&=&\frac{5}{2}\delta n_1+12 \delta n_2 \geq 0\,,
\end{eqnarray}
with $T^{bulk}_{00}+T^{bulk}_{11}=0$. These simplify to $n_1\leq 1/2$ and $n_2\geq -1/24$.  Thus the region in fig.1 that respects the null energy condition is smaller than that allowed by the positivity of relative entropy.

For a general constant stress tensor in general $d$ we proceed as follows. We note that for a metric of the form in eq.(\ref{fgmetric}), with $g_{\mu\nu}$ a function of $z$ only, we have \cite{headrickcomp}
\begin{eqnarray}
R_{\mu\nu}&=&R'_{\mu\nu}-(z\partial_z K_{\mu\nu}+ K K_{\mu\nu}-2 K_{\mu\kappa}K^{\kappa}_{\nu})\,,\\
R_{\mu z} &=&0\,,\\
z^2 R_{zz} &=& - g^{\mu\nu}z \partial_z K_{\mu\nu}+K^{\mu\nu}K_{\mu\nu}\,,\\
R&=& R'-(2  z g^{\mu\nu} \partial_z K_{\mu\nu}+ K^2-3 K^{\mu\nu}K_{\mu\nu})\,,
 \end{eqnarray}
where $K_{\mu\nu}=\frac{1}{2} z\partial_z g_{\mu\nu}$. Here $'$ denotes a quantity computed with $g_{\mu\nu}$.
Using these it is straightforward (but tedious) to compute (setting $L=1$ for convenience, defining $S_{\mu\nu}=n_1 T_{\mu\alpha}T^{\alpha}_\nu+n_2 \eta_{\mu\nu} T_{\alpha\beta}T^{\alpha\beta}$ and aborbing the factors of $a$ into $T_{\mu\nu}$; the raising and lowering of indices on $T_{\mu\nu}, S_{\mu\nu}$ are done with $\eta_{\mu\nu}.$ Also we have used $T_\mu^\mu=0$.)
\begin{eqnarray}
g^{\mu\nu} &=&z^2[\eta^{\mu\nu}-T^{\mu\nu}z^d+(T^{\mu\alpha}T_\alpha^\nu-S^{\mu\nu})z^{2d}]\,, \\
K_{\mu\nu} &=&-\frac{1}{2z^2} (2\eta_{\mu\nu}-(d-2) z^d T_{\mu\nu}-2(d-1)z^{2d}S_{\mu\nu})\,,\\
K_\mu^\nu &=&-\frac{1}{2}[2\delta_\mu^\nu-d z^d T_\mu^\nu+d z^{2d}(T_{\mu\alpha}T^{\alpha\nu}-2 S_\mu^\nu)]\,,\\
K &=& -\frac{1}{2} [2 d+d z^{2d}(T_{\alpha\beta}T^{\alpha\beta}- 2 S_\alpha^\alpha)]\,,\\
K^{\mu\nu} &=& -\frac{z^2}{2}[2\eta^{\mu\nu}-(d+2) z^d T^{\mu\nu}+2(d+1)z^{2d}(T^\mu_\alpha T^{\nu\alpha}-S^{\mu\nu})]\,,\\
z g^{\mu\nu} \partial_z K_{\mu\nu}&=&\frac{1}{2}[4 d+z^{2d}(4d(d-2) S_\alpha^\alpha-d(d-4) T_{\alpha\beta} T^{\alpha\beta} )]\,,\\ 
K_{\mu\nu}K^{\mu\nu} &=& d+\frac{z^{2d}}{4}[d(d+4) T_{\alpha\beta}T^{\alpha\beta}-8 d S_\alpha^\alpha]\,,\\
 K_{\mu\kappa}K^{\kappa}_{\nu}&=& \frac{1}{4z^2}[4\eta_{\mu\nu}-4(d-1)z^d T_{\mu\nu} +z^{2d}(d^2 T_\mu^\kappa T_{\kappa \nu}-4(2d-1) S_{\mu\nu})]\,.
\end{eqnarray}
Using these we find
\begin{eqnarray}
T^{bulk}_{zz} &=&-d(d-1) z^{2d-2}T_{\alpha\beta}T^{\alpha\beta} (\delta n_1+d \delta n_2)\,,\\
T^{bulk}_{\mu\nu} &=& d^2 z^{2d-2}\left[-\delta n_1 T_{\mu \kappa}T^\kappa_\nu+\eta_{\mu\nu} T_{\alpha\beta}T^{\alpha\beta}(\delta n_1+(d-1)\delta n_2)\right]\,.
\end{eqnarray}
Here $\delta n_1=n_1-1/2$ and $\delta n_2=n_2+1/(8(d-1))$ i.e., the deviations from the Einstein values.
Now we are in a position to ask if the matter supporting this bulk stress tensor satisfies the null energy conditions or not. First we note that $T^{bulk}_{00}+T^{bulk}_{11}\geq 0$ immediately leads to 
\be
-d^2 (-T_{00}^2+T_{ij}^2)\delta n_1 \geq 0\,.
\ee
This leads to a definite sign for $\delta n_1$ if and only if $ (-T_{00}^2+T_{ij}^2)$ has a definite sign. But in general, there is no reason for this combination to have a definite sign. So we are led to suspect that for a generic stress tensor, $\delta n_1=0$. To confirm this suspicion let us look at $T^{bulk}_{zz}+T^{bulk}_{00}$. 
\begin{eqnarray}
&& T^{bulk}_{zz}+T^{bulk}_{00}= \nonumber \\ &&-d\left[(d-1)T_{00}^2 (\delta n_1+2 d \delta n_2)+ T_{ij}^2 [(2d-1)\delta n_1+2 d(d-1)\delta n_2]+ T_{0i}^2 [(2-3d)\delta n_1-4d(d-1)\delta n_2]\right]\,.\nonumber\\
\end{eqnarray}
As in the relative entropy analysis, we write the RHS as $V^T M V$ where $V$ is a $(d-1)(d+2)/2$ dimensional vector whose non-zero independent components are the $T_{00}, T_{ij}, T_{0i}$'s. Then we demand that the eigenvalues of $M$ are positive for the null energy condition to hold for a generic constant traceless stress tensor $T_{\mu\nu}$. This yields
\begin{eqnarray}
(3d-2)\delta n_1+4d(d-1)\delta n_2 &\geq &0 \,,\\
(2d-1)\delta n_1+2d(d-1)\delta n_2 &\leq &0 \,,\\
\delta n_1+2(d-1)\delta n_2 &\leq & 0\,.
\end{eqnarray}
Only for $\delta n_1=\delta n_2=0$ are these inequalities satisfied for $d>2$. Thus the null energy condition picks out the Einstein value if we ask if for a generic constant stress tensor the $O(T^2)$ terms are supported by matter. Of course as we saw for $d=4$ we can turn on $T_{0i}$ and set everything else to zero, there would be a region in the $n_1,n_2$ parameter space where the null energy condition and the positivity of the relative entropy would hold (this corresponds to the region between the red and blue solid lines in fig.1). For the generic case, only the Einstein value is picked out. To emphasise, that the Einstein value was picked out for the generic case, relied only on the null energy condition analysis and did not rely on the positivity of the relative entropy. To summarize, we found that there exists a larger class of theories in the $(n_1, n_2)$ parameter space than just the Einstein theory. However, except at the Einstein point, we found that there always exists some matter stress tensor which violates the bulk null energy condition.

\section{Relative entropy in Gauss-Bonnet holography}
In this section we will calculate relative entropy for excited states in Gauss-Bonnet gravity. For definiteness, we will consider $d=4$ or 5-dimensional bulk. We will follow the conventions in \cite{gbholo}. 
The total action is given by
\be 
I=I_{bulk}+I_{GH}+I_{ct}\,,
\ee
where
\be
I_{bulk}=\int d^5 x \sqrt{g}\left[R+\frac{12}{L^2}+\frac{\lambda}{2}L^2 (R_{ABCD}R^{ABCD}-4R_{AB}R^{AB}+R^2)\right]\,.
\ee
The generalized Gibbons-Hawking term is given by \cite{kcube}
\be\label{surface1}
I_{GH}=- \frac{1}{\lp^{3}} \int d^{4}x \sqrt{\gamma}\big[K-\lambda L^{2}(2G_{\mu\nu}K^{\mu\nu}+\frac{1}{3}(K^{3}-3KK_{2}+2K_{3})\big]\,.\\
\ee
Here $G_{\mu\nu}=R_{\mu\nu}-1/2 \gamma_{\mu\nu} R$ made from the boundary $\gamma_{\mu\nu}$, $K_{2}=K_{\mu\nu}K^{\mu\nu}$ and $K_{3}=K^{\alpha}_{\b}K^{\b}_{\g}K^{\g}_{\a}$. $K_{\mu\nu}$ is the extrinsic curvature and $K=K_\a^\a$.
The counterterm action  $I_{ct}$ is needed for the cancellation of the power law divergences in $I_{tot}$. For our case this works out to be \cite{yale,bhss} ($\tilde L$ and $f_\infty$ are defined below)
\be
I_{ct}=\frac{1}{\lp^{3}} \int d^{4}x\, \sqrt{\gamma}\big[ c_1\frac{3}{\tilde L}+ c_2 \frac{\tilde L}{4}\hat R\big]\,,
\ee
where $\hat R$ is the  four dimensional Ricci scalar and $ c_1=1-\frac{2}{3}f_{\infty}\lambda $ and $c_2=1+2f_{\infty}\lambda\,.$

The equations of motion are given by
\be\label{gbeq}
R_{AB}-\frac{1}{2} g_{AB}  R-\frac{6}{L^2}g_{AB}-\frac{\lambda L^2}{2} H_{AB}=0,
\ee
where
$$
 H_{AB}=\frac{1}{2}g_{AB}(R^2-4R_{MN}R^{MN}+R_{MNRS}R^{MNRS})-2 RR_{AB}+4R_A^R R_{RB} -2R_{AMNS}{R_B}^{MNS}-4R^{MN}R_{MABN} \,.
$$
AdS$_5$ given by
\be
ds^2=\frac{\tilde{L}^2}{z^2}\left(dz^2-dt^2+dx_1^2+dx_2^2+dx_3^2\right)
\ee
where $\tilde L=L/\sqrt{f_\infty}$ with $1-f_\infty+\lambda f_\infty^2=0$. The dual CFT is characterized by the central charges $c,a$ appearing in the trace anomaly\cite{gbholo, myersme}:
\be
c=\frac{\pi^2 \tilde L^3}{\lp^3}(1-2\lambda f_\infty)\,,\quad a=\frac{\pi^2 \tilde L^3}{\lp^3}(1-6\lambda f_\infty)\,.
\ee
The CFT stress tensor two point function is given by
\be
\langle T_{\mu\nu}(x) T_{\rho\sigma}(0)\rangle=\frac{40 c}{\pi^2(x^2)^4} {\mathcal I}_{\mu\nu,\rho\sigma}(x)\,,
\ee
where $\mathcal{I}$ is a function of $x$ and the positivity of the two point function leads to $c>0$.

We will need the formula for the holographic stress tensor (see eg.\cite{Liu:2008zf}) 
\be\label{holT}
T_{\mu\nu}=\frac{1}{\ell_{p}^{3}}[K_{\mu\nu}-g_{\mu\nu}K+\lambda L^{2} (q_{\mu\nu}-\frac{1}{3}g_{\mu\nu}q)]-\frac{3}{\tilde L} c_1 \gamma_{\mu\nu}+\frac{\tilde L}{2} c_2 [R_{\mu\nu}(\gamma)-\frac{1}{2} \gamma_{\mu\nu}R(\gamma)]\,,
\ee
where \\
$ q=h^{\mu\nu}q_{\mu\nu}\,$ \\ $ q_{\mu\nu}=2 K K_{\mu\alpha}K^{\alpha}_{\nu} -2K_{\mu\alpha} K^{\alpha\beta}K_{\beta\nu}+K_{\mu\nu}(K_{\alpha\beta}K^{\alpha\beta}-K^{2})+2K R_{\mu\nu}+RK_{\mu\nu}-2K^{\alpha\beta}R_{\alpha\mu\nu\beta}-4R_{(\mu}^{\alpha}K_{\nu)\alpha}\,. $ 
The terms proportional to $c_1,c_2$ come from $I_{ct}$.

We also note that the GB coupling $\lambda$ is bounded. Following \cite{hofman} for the calculation of the three point correlation function of stress tensor one needs to compute a energy flux which comes form the insertion of $\epsilon_{ij}T_{ij}\,,$ where $\epsilon_{ij}$ and $T_{ij}$ are the polarization tensor and stress tensor respectively. Demanding the positivity of this energy flux in the holographic set up we get the following three constraints and from those we obtain bounds on $\lambda\,.$ These coincide with the bounds arising from micro-causality \cite {micro1}.
\begin{align}
\begin{split}\label{lamcon}
\rm{Tensor\,channel:}&\,\, 1- 10f_{\infty}\lambda\geq 0 \Rightarrow \lambda \leq \frac{9}{100}\\
\rm{Vector\,channel:}&\,\, 1+2f_{\infty}\lambda\geq 0 \Rightarrow -\frac{3}{4}\leq \lambda\leq\,\frac{1}{4}\\
\rm{Scalar\,channel:}&\,\, 1+6f_{\infty}\lambda\geq 0 \Rightarrow -\frac{7}{36}\leq \lambda\leq\,\frac{1}{4}\\
\end{split}
\end{align}
From this we get $$ -\frac{7}{36} \leq \lambda \leq \frac{9}{100}\,.$$
This is the same as the condition $1-4 f_{\infty}\lambda -60 f_{\infty}^{2}\lambda^{2}\geq 0$. 

\subsection{Linear order calculations}
We are interested in considering the excited state to be a perturbative excitation of the ground state. At linear order in the perturbation $\Delta H=\Delta S$. Let us review the argument \cite{Relative} why.
Let $\rho_0$ be a reference state. Now let $\rho(\alpha)$ be a continuous family of states dependent on a parameter $\alpha$ that runs over all possible values. We choose the parametrization such that $\rho(\alpha=0)=\rho_0$. Now relative entropy vanishes for two states that are equal. So we must have $S(\rho(0)|\rho_0)=0$ and also $S(\rho(\alpha\rightarrow\epsilon\pm)|\rho_0)\rightarrow 0+>0$ where $\epsilon$ is a small positive valued number denoting a small perturbation from the reference state $\rho_0$. This means at $\alpha=0$ we must have,\, $d(S(\rho(\alpha)|\rho_0))/d\alpha=0$. Or equivalently at the linear order of the perturbation $\epsilon$,
\be\label{equality}
\Delta H=\Delta S\,,
\ee
which follows from eq.(\ref{dhds}).
We can demonstrate this with a simple example\footnote{The change in entanglement entropy for excited states in GB holography has been considered in \cite{chinese1}.}. Let $\rho_0$ to be the vacuum of the CFT$_4$ whose holographic dual is the empty AdS$_5$ (our linearized results are a sub-case of the more general case worked out in \cite{einstein3}),
\be
ds^2=\frac{\tilde{L}^2}{z^2}\left(dz^2-dt^2+dx_1^2+dx_2^2+dx_3^2\right)
\ee
We choose $\rho_1$ to be the dual of a metric which is being perturbed around the empty AdS. Following \cite{Relative}, we take the perturbation to be of the form,
\be
\delta g_{\mu\nu}=\frac{\lp^{3}}{2\tilde{L}^{3}}z^2\sum_n z^{2n}T^{(n)}_{\mu\nu}\,.
\ee
To keep track of the perturbation we keep the components of $T^{(n)}_{\mu\nu}$ proportional to a small number $\epsilon$. We compute the entanglement entropy from the Jacobson-Myers functional,
\be\label{JM}
S=\frac{2\pi}{\lp^3}\int d^3 x \sqrt{h}(1+\lambda L^2 \R)+ \frac{4\pi}{\lp^3}\lambda L^2\int d^2x \sqrt{h}\,\,\K\,.
\ee
Here, $h_{ab}$ is the induced metric on the minimal surface and $\R$ and $\K$ are respectively the intrinsic ricci scalar and extrinsic curvature evaluated on that surface. To simplify notation, we will set $L=1\,.$ 
 The minimal surface equation is given by
\be \label{condition1}
\K+\lambda L^{2}(\R\K-2\R_{ij}\K^{ij})=0\,,
\ee
which was derived in \cite{ab, Chen} following \cite{mukund}.
For the spherical entangling surface in the unperturbed metric the following continues to be an exact solution 
\be
z=z_0=\sqrt{R^2-r^2}\,.
\ee
In the perturbed case, it  changes to 
\be
z=z_0+\epsilon\, z_1\,.
\ee
However note that we obtained $z_0$ by extremization.  Hence $z_1$ can only contribute to a quadratic order in $\epsilon$ and not at linear order. Thus at linear order we can set $z_1=0$.
Using $z=z_0$ to compute (\ref{JM}) and then extracting the terms proportional to $\epsilon$ gives us $\Delta S$. 
Now we can calculate the modular hamiltonian, from the formula in eq.(\ref{modH}),
where $T_{00}$ is obtained in holography using eq.(\ref{holT}).
Since $T_{00}=0$, for empty AdS, this directly gives $\Delta H$. Now we will demonstrate the equality in eq.(\ref{equality}) by considering a special case (we have checked that this holds in the other examples considered below as well).

Using Gauss-Bonnet eom, we can determine $T^{(n)}_{\mu\nu}$ in terms of the lowest mode $T^{(0)}_{\mu\nu}$. It turns out they are all derivatives of $T^{(0)}_{\mu\nu}$. To keep it simple we take $T^{(0)}_{\mu\nu}$ to be a constant. Also note that to satisfy GB eom, we must have traceless and divergenceless conditions on $T^{(0)}_{\mu\nu}$,
\be\label{conditions}
{T^{(0)}}^\mu_\mu=0\hspace{1cm}\mbox{and}\hspace{1cm}\partial_\mu {T^{(0)}}^\mu_\nu=0\,.
\ee
Consider an isotropic perturbation
\be
{T^{(0)}}_{\mu\nu}=\left(\mathcal{E},\frac{\mathcal{E}}{3},\frac{\mathcal{E}}{3},\frac{\mathcal{E}}{3}\right)
\ee
Note that this satisfies the conditions in eq.(\ref{conditions}). However the holographic dual tensor $T_{\mu\nu}$ is not same as ${T^{(0)}}_{\mu\nu}$. We compute it from eq.(\ref{holT}) as,
\be
T_{\mu\nu}=(1-2f_{\infty}\lambda)\left(\mathcal{E},\frac{\mathcal{E}}{3},\frac{\mathcal{E}}{3},\frac{\mathcal{E}}{3}\right)
\ee
Now using (\ref{modH}) one gets\footnote{There is a typo in eq.(6.29) in \cite{einstein3} for $\langle T_{\mu\nu}\rangle$. There is a factor of 2 missing in front of the term proportional to $a_1$ in that expression. Taking this into account our expression agrees with their both for GB and for the general $R^2$ theory discussed in appendix C.},
$$
\Delta H=\frac{8\pi^2 \tilde L^3 \mathcal{E} R^4}{15 \lp^3}(1-2f_\infty\lambda)\,.
$$
As discussed before, we can compute $\Delta S$ from (\ref{JM}) with $z=\sqrt{R^2-r^2}$, and then take out the $\epsilon$ order coefficients. We obtain,
\be
\sqrt{h}(1+\lambda L^{2} \R)=-\frac{\mathcal{E}  \left(R^2 \left(3+30 f_\infty  \lambda \right)-r^2 \left(1+58 f_\infty  \lambda \right)\right)}{6 f_\infty^{3/2} R}
\ee
from which we calculate,
\be
\Delta S = \frac{8\pi^2 \tilde L^3 \mathcal{E} R^4}{15 \lp^3}(1-2f_\infty\lambda)\,.
\ee
This demonstrates $\Delta H=\Delta S$ for an isotropic perturbation.


\subsection{Quadratic corrections}
Now we turn to the more interesting case of quadratic corrections which lead to inequalities. We take the following form for the boundary metric,
\be\label{metricpert2}
z^2 g_{\mu\nu}=\eta_{\mu\nu}+z^d T_{\mu\nu} + z^{2d}(n_1T_{\mu\alpha}T_\nu^\alpha+n_2\ \eta_{\mu\nu}T_{\alpha\beta}T^{\alpha\beta})+\cdots
\ee
where compared to eq.(\ref{metricpert}) we have absorbed a factor of $a$ into the stress tensor. We need to fix the numbers $n_1$ and $n_2$. By plugging into the GB equations of motion given by eq.(\ref{gbeq}), we find that
\be
-3 (n_1 + 4 n_2) + f_\infty (1 + 6 n_1 \lambda + 24 n_2 \lambda)=0
\ee
\be
n_1 (9 - 17 f_\infty + 25 f_\infty^2 \lambda) - 
  4 (4 f_\infty^3 \lambda - 3 n_2 (1 - 9 f_\infty + 17 f_\infty^2 \lambda))=0
  \ee
Solving the two equations and using the relation $1-f_\infty+f_\infty^2\lambda=0$ we get 
\be
n_1=\frac{1}{2}\frac{1+2f_{\infty}\lambda}{1-2f_{\infty \lambda}}\hspace{1 cm}\mbox{and}\hspace{1 cm}n_2=-\frac{1}{24}\frac{1+6f_{\infty}\lambda}{1-2 f_{\infty}\lambda}\,.
\ee
These results match  with the $\lambda=0$ case given in  \cite{Relative}\footnote{Notice a curious fact. If we demanded that $n_1\geq 0$ and $n_2 \leq 0$, or in other words even in GB gravity they have the same sign as in Einstein gravity then with $c>0$, we would get 
$$
1+2f_{\infty}\lambda\geq 0\,,\quad 1+6f_{\infty}\lambda\geq 0\,.
$$
But these are nothing but the scalar and vector channel constraints in eq.(\ref{lamcon})! These leads us to wonder if entanglement entropy knows about the causality constraints. }.
\subsection{Constant $T_{\mu\nu}$}
The next step is to calculate the second order change in $\Delta S$. For a general but  constant stress tensor we can guess the following form of the second order correction of entropy from Lorentz invariance,
\be\label{geneq}
\Delta^{(2)} S= C_1 T^2 +C_2 T_{ij}T^{ij} +C_3 T_{0i}T^{0i}
\ee
where $T$ denotes the trace of the spatial part of the stress tensor $T_{\mu\nu}$. The latin indices run from 1 to 3, and denote the spatial part of a tensor. They are raised with $\eta_{ij}$. 
Our task is to identify the constants $C_i$'s for a non-zero $\lambda$. The only condition on the stress tensor is that it is  symmetric and traceless. To do the perturbative analysis we assume that the components of the stress tensor are  proportional to a perturbative parameter $\epsilon$.  Also we have absorbed a parameter $a$ in the stress tensor. The background metric will be changed in the quadratic order as given in (\ref{metricpert}). Now, assume that the minimal surface $z_0=\sqrt{R^2-r^2}$ is modified as
\be
z=z_0+ \epsilon z_1\,.
\ee
 $z_1$ contributes at the quadratic order in  the JM functional (\ref{JM}).  So it is sufficient to consider only the  first order fluctuation to the entangling surface. 
Next we expand the entropy functional  upto quadratic order and then extract the terms proportional to $\epsilon^2$ which gives the quadratic correction to the entropy.  We vary it with respect to $z_1$. This gives us the equation of motion for $z_1$. We find the solution and put it back to $\Delta^{(2)} S$. Since it was shown that at linear order, $\Delta S=\Delta H$, at second order we must have $\Delta^{(2)} S>0$. 
We get the following equation for $z_{1}\,,$
\be
(1-2f_\infty \lambda)\left[\partial^2(z_0 z_1)-\frac{x^i x^j}{R^2}\partial_{i}\partial_{j}(z_0z_1)-(R^2-r^2)^2\left(T+3T_x\right)\right]=0\,,
\ee
with the solution,
\be
z_1=-\frac{ R^2z_0^3}{10}\left(T+T_x \right)\,.
\ee
Notice that the equation is the same as what appears in the Einstein case upto the overall factor of $(1-2\lambda f_\infty)$. 
 The Gibbons Hawking term doesn't contribute to the action when we put in the solution. Alternatively, we could have taken the action and integrated all terms involving $z_1''(x)$'s by part and cast it in the conventional form. The surface term resulting from this will cancel with the appropriate Gibbons-Hawking term. We have checked both approaches and have got the same result. Integrating the resulting action over the volume of the entangling region, we obtain the second order correction to the entropy,
\be
\Delta^{(2)} S=-\frac{8\pi^{3} \tilde L^{3}(1-2f_{\infty}\lambda)}{\lp^{3} }\left(
C_1T^2+C_2T_{ij}^2+C_3T_{i0}^2 \right)\,,
\ee
$C_{1},C_{2},C_{3}$ are same as the Einstein values obtained in section 2. 

Note that this is just a factor of $(1-2f_{\infty}\lambda)$ times what is obtained in the Einstein gravity (the Einstein result was manifestly negative).  This can be cross-checked easily on a computer by suitably turning on various components  of  the stress tensor  and identifying various tensor structures. \par 
Now from the discussions in the previous sections, it is clear that this quantity has to be negative. The only constraint to ensure $\Delta^{(2)} S<0$ is
\be
1-2f_{\infty}\lambda>0\,.
\ee
This is equivalent to saying the central charge $c>0$ which also is the condition needed for the positivity of the two point function of the field theory stress tensor. The condition $\lambda < 1/4$ ensures that this holds. If this inequality on $\lambda$ did not hold, the corresponding vacuum would have ghosts \cite{gbholo}.

\subsection{Shockwave background}
Up to this point we have only considered constant stress-tensor. It is interesting to ask if we get non-trivial constraints for $T_{\mu\nu}$ not constant. To explore a nontrivial case of non-constant $T_{\mu\nu}$, consider the following 5 dimensional metric
\be\label{new}
ds^2=\frac{\tilde{L}^2}{z^2}(dz^2+dx_{\mu}dx^{\mu}+f(t+x_3)W(z,x_1,x_2)(dt+dx_3)^2)
\ee
where $\mu={1,2}$.

The above metric solves the GB equation exactly, given that $W(z,x_1,x_2)$ satisfies the following differential equation,
\be\label{Weqn}
\partial_z^2 W +\partial_{x_1}^2 W +\partial_{x_2}^2 W=-\frac{3}{z}\partial_z W \,,
\ee
with no constraint on $f(t+x_3)$. If $f=\delta(t+x_3)$ then this is the shockwave metric considered for example in \cite{hofman} to derive constraints on higher derivative gravity theories. We will set $f=1$ and in a slight abuse of terminology continue to refer the metric as a shockwave.
 $W(z,x_1,x_2)$ is taken as
\be \label{sch}
W(z,x_1,x_2)=\frac{ \tilde{L}^2 z^4}{(z^2+(x_1-x_1')^2+(x_2-x_2')^2)^3}\,.
\ee
Here $(x_1',x_2')$ represent the point where the disturbance is peaked. Since in our calculations we perturb the background metric, we should choose $x_1'$ and $x_2'$ to be outside the entangling region. With this in mind we  proceed with the second order calculation. 
 Next we consider a shockwave disturbance localised just outside the entangling surface. We will set $x_2'=0$   in (\ref{sch}). We start with the following metric which is obtained by expanding $W$ around $z=0$ and retaining the first two terms in the expansion,
\be 
ds^2=\frac{\tilde {L}^2}{z^2}(dz^2+dx_{\mu}dx^{\mu}+(\frac{ z^{4}\tilde{L}^2\epsilon^{3}}{(x_{1}^{2}+(x_{2}-  x_2')^{2})^{3}}-\frac{3 z^{6} \tilde{L}^2\epsilon^{4} }{(x_{1}^{2}+(x_{2}-  x_2')^{2})^{4}})(dt+dx_3)^2)
\ee
The $\epsilon$ factors have been inserted to keep track of the order of the expansion and matches with the power appear in the denominator. If we write the entangling surface as $z=z_0+\epsilon^3 z_1$ then the quadratic terms in $z_1$ will involve $\epsilon^6$ which is at a higher order than the second order term in the metric above. Thus we expect to see an inequality $\Delta H>\Delta S$ with the above metric setting $z_1=0$.
We thus evaluate the entropy functional considering only the unperturbed entangling surface  and expand it upto $\epsilon^{4}$ and pick out the $\epsilon^{4}$ term which gives the first  leading order change in the relative entropy.  The integrand is shown below,
\begin{align}
\begin{split}
\Delta^{(2)} S &=\frac{2\pi}{\lp^{3}}\int dx_{3} dx_{1} dx_2 \frac{3 L^{5}}{2 R f_{\infty}^{5/2} (x_3^{2}+(x_2-x_2')^{2})^6}\big[(x_3^{2}+x_2^{2}+x_1^{2}-R^{2})(40f_{\infty}(x_3^{2}+x_2^{2}+x_1^{2}-R^{2})\\ &(4 R^2(x_3^2+(x_2-x_2')^2)-4(x_3^4+x_3^2(x_1^2+2x_2(x_2-x_2'))+(x_1^2+x_2^2)(x_2-x_2')^2))\lambda+16 f_{\infty}\\&(R^2-x_3^2-x_2^2-x_1^2)(x_3^2+(x_2-x_2')^2)(2R^2-13x_3^2-2x_1^2-13x_2^2+12x_2x_2')\lambda\\&-(x_3^2+(x_2-x_2')^2)^2(60f_{\infty}(x_3^2+x-2^2)\lambda-R^2(1+18f_{\infty}\lambda)+x_1^2(1+18f_{\infty}\lambda)))\,.\\
\end{split}
\end{align}
Then we perform the integration over $x_3$ which goes from $-\sqrt{R^{2}-r^2}$ to $\sqrt{R^2-r^2}$ and 
$ x_{1}= r \cos(\theta)\,\,, \,\,x_{2}= r \sin(\theta)\,.$ Now after some algebraic manipulation we can write the integrand as,
\be
\Delta^{(2)}S=\frac{2\pi L^{5}}{2\lp^{3} f_{\infty}^{5/2} R (r^2+x_2'^2-2 r x_2' \sin(\theta ))^6}(f_1+f_2 \sin(\theta )+f_3 \sin(\theta )^2)
\ee
where $f_{1},f_{2},f_{3}$ are some function of $r$ and $\lambda\,.$
 Integral over $\theta$  goes from $0 \,\,\rm{to}\,\, 2\pi$ and integral over $r$ goes from $0$ to $R.$  We first perform the $\theta$ integral. To perform the $\theta$ integral  we have used the following integral identity: $$ \int_{0}^{2\pi} \frac{d\theta}{a+b \sin(\theta)}=\frac{2\pi}{\sqrt{a^{2}-b^{2}}}\,,$$  Finally we get,
\begin{align}
\begin{split}
\Delta^{(2)}S=\frac{2\pi}{\lp^3}\int_{0}^{R} dr \big[-\frac{L^{5}}{240f_{\infty}^{5/2}R}& \big (f_1 (-\frac{30 \left(8 a^5+40 a^3 b^2+15 a b^4\right) \pi }{\left(a^2-b^2\right)^{11/2}})+f_2 (\frac{90 b \left(8 a^4+12 a^2 b^2+b^4\right) \pi }{\left(a^2-b^2\right)^{11/2}})\\& f_{3} (-\frac{30 \left(4 a^5+41 a^3 b^2+18 a b^4\right) \pi }{\left(a^2-b^2\right)^{11/2}})\big)\big]\,,\\
\end{split}
\end{align}
where, $a^2= r^2+x_2'^2$ and $ b=-2 r x_2'\,.$ Next we perform the $r$ integration. The leading contribution in $\Delta^{(2)} S $ comes form the lower limit of the $r$ integral which is shown below. 
\be
\Delta^{(2)} S= \frac{\pi^2 L^{5}}{96 \lp^3 f_{\infty}^{5/2} R^2}(1-2f_{\infty}\lambda) f(x_2')\,,\\
\ee
where, $f(x_2')$ is a negative valued function given by
$$ f(x_2')= \frac{  \left(\sqrt{x_2'^2-1} \left(-136+72 x_2'^2-56 x_2'^4+15 x_2'^6\right)-3 \left(32-16 x_2'^2+36 x_2'^4-22 x_2'^6+5 x_2'^8\right) Csc^{-1}(x_2')\right)}{ \left(x_2'^2-1\right)^{9/2}}\,,$$
and plotted in fig.2.
To satisfy, $ \Delta S\leq \Delta H$ we will get, $1-2f_{\infty}\lambda \geq 0$ or in other words $c>0$. Note that in order for us to be able to expand in small $z$, the perturbation needs to be located far away from the entangling surface. This is because in the denominator in $W$ we had $z^2+x_1^2+(x_2-x_2')^2$. When we plug in $z=z_0$, the maximum value for $z$ is $R$ and this happens when $x_1=x_2=0$. Thus we will need $R\ll x_2'$ for the expansion to be valid. It will be interesting to see what happens as we move the perturbation closer and closer to the entangling surface. However this appears to be a very hard problem.
\begin{figure}[ht]
\centering\includegraphics[scale=0.75 ]{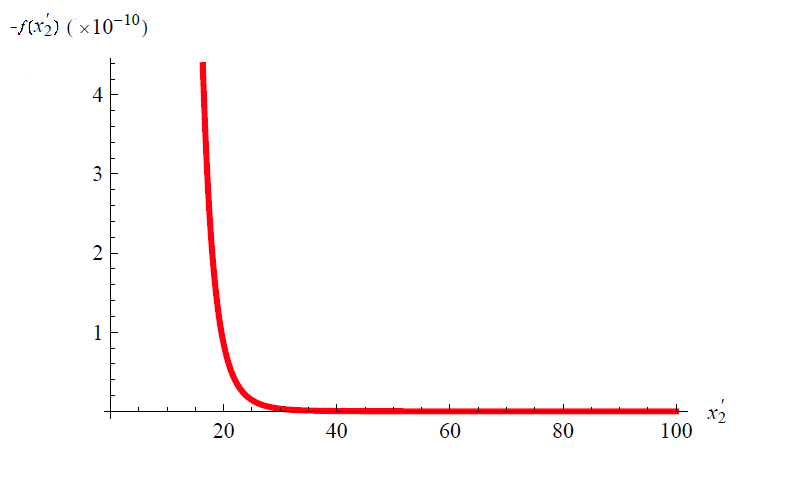}
\caption{Negative of the function $f(x_2')$ is plotted which is a positive valued function}
\end{figure}

\subsection{Correction from additional operators}
In this section we consider perturbed states in which certain additional operators acquire nontrivial vacuum expectation value. Our analysis will follow \cite{Relative}. The holographic dual of these operators will involve additional massive fields in the bulk. We will show that even for such cases in Gauss-Bonnet gravity, the relation $\Delta H>\Delta S$ will hold. Again we are in AdS$_5$ with the bulk action given by,
\be
I=\int d^5x\sqrt{-G}\left(R+\frac{12}{L^2}+\frac{\lambda L^2}{2}\left(R^2-4R_{AB}^2+R_{ABCD}^2\right)-\frac{1}{2}(\partial\phi)^2-\frac{1}{2}m^2\phi^2\right)\,,
\ee
where we have added a massive scalar field which acts as a bulk dual of a scalar operator of dimension $\Delta$. When $m^2=\Delta(4-\Delta)$, the field  $\phi$ behaves asymptotically as,
\be
\phi=\gamma\mathcal{O}z^\Delta\,.
\ee
Now we can work out the stress tensor corresponding to this from the formula,
\be
T_{AB}=\frac{1}{2}\partial_A\phi\partial_B\phi-\frac{1}{4}g_{AB}((\partial\phi)^2+m^2\phi^2)\,.
\ee
This will result in the following change to the boundary metric boundary metric,
\be\label{scalarpert}
z^2\delta g_{\mu\nu}=az^d\sum_n z^{2n} T^{(n)}_{\mu\nu}+z^{2\Delta}\sum_n z^{2n}\sigma_{\mu\nu}^{(n)}
\ee
where we must have,
\be
\sigma_{\mu\nu}^{(0)}=-\frac{\gamma^2}{12(1-2f_{\infty}\lambda)}\eta_{\mu\nu}\mathcal{O}^2\,.
\ee
in order to satisfy Gauss-Bonnet eom. The higher modes, namely $\sigma^{(n)}_{\mu\nu}\,\,(n>0)$ are composed of derivatives of $\sigma^{(0)}_{\mu\nu}$. As in \cite{Relative}, we consider $\mathcal{O}$ to be slowly varying and, hence, neglect the higher modes.

It is not necessary to find any correction to the entangling surface. There are two different perturbations, both in their first orders, and using the $z_0$ minimal surface to compute $\Delta S$ will suffice. The correction to entropy will have two parts,
\be
\Delta S=\Delta_T S+\Delta_\mathcal{O} S\,.
\ee
The first part, $\Delta_TS$ comes from the holographic boundary stress tensor $T_{\mu\nu}$, and its the same as what we calculated before for the linear order. The second part comes from the scalar field and is obtained by calculating the area functional with the metric of eq.(\ref{scalarpert}).
\be
\Delta_\mathcal{O}S=-\frac{\pi ^{3/2} R^{2 \Delta } \gamma ^2 (-2+3 \Delta ) \Gamma [-1+\Delta ] \Omega _{d-2}}{48 \,a\,  \Gamma \left[\frac{1}{2}+\Delta \right]}\mathcal{O}^2\,.
\ee
Note that the result is independent of $\lambda$. Since the result is negative it seems the metric already knows of the positivity of relative entropy even for Gauss-Bonnet provided the unitarity bounds are respected. 

\section{Relative entropy for an anisotropic plasma}
 We now want to turn our attention to a holographic anisotropic plasma--there is going to be a surprise in store. We consider the holographic dual of the deformed ${\mathcal{N}}=4$ SYM where the deformation is generated by anisotropy along one spatial direction \text{viz.} 
\be 
S=S_{{\mathcal{N}}=4}+\frac{1}{8\pi^2}\int \theta(z) \  \text{Tr} \ F\wedge F,
\ee
$\theta$ is the field generating anisotropy along the $z$ direction. The holographic dual is the Einstein-dilaton-axion system given by
\be 
S_{bulk}=\frac{1}{2\lp^3}\int_{{\mathcal{M}}}\sqrt{-g}(R+\frac{12}{L^2}-\frac{1}{2}(\partial\phi)^2-\frac{1}{2}e^{2\phi}(\partial\chi)^2)+\frac{1}{2\lp^3}\int_{\partial{\mathcal{M}}}\sqrt{-\g}2K,
\ee
where $\phi$ is the dilaton and at the level of the solution is taken to be a function of the AdS radius only and $\chi$ is the axion dual to the gauge theory $\theta$-term, responsible for inducing anisotropy, which is taken to be $\chi=\rho x_3$. This model was proposed and studied in detail in \cite{mt12}. The low anisotropy regime corresponding to $\rho/T \ll 1$ in this model is unstable \cite{mt12}.

\noindent The metric equations are given by ($L=1$)
\be 
R_{MN}-\frac{1}{2}R g_{MN}-6 g_{MN}=T_{MN},
\ee 
where the bulk matter stress tensor is given as
\be \label{anT}
T_{MN}=\frac{1}{2}\partial_{M}\phi\partial_{N}\phi-\frac{1}{4}(\partial\phi)^2 g_{MN}+\frac{1}{2}e^{2\phi}\partial_{M}\chi\partial_{N}\chi-\frac{1}{4}e^{2\phi}(\partial\chi)^2 g_{MN}\,.
\ee

\noindent The metric, $\phi$ and $\chi$ equations can be written as
\begin{align}
\begin{split}
R_{MN}+4 g_{MN}-\frac{1}{2}\partial_{M}\phi\partial_{N}\phi-\frac{1}{2}e^{2\phi}\partial_{M}\chi\partial_{N}\chi=0,\\
\nabla^2\phi-e^{2\phi}(\partial\chi)^2=0,\\
\nabla^2\chi=0\,.
\end{split}
\end{align}
%

\par

 \noindent The metric in the FG coordinates is given by
\be 
ds^2=\frac{dz^2}{z^2}+\frac{1}{z^2}\g_{\m\n}(z, x^{i})dx^{\m}dx^{\n},
\ee
where
\begin{align}
\begin{split}
\g_{tt}&=-1+\frac{\rho^{2}}{24}z^{2}+\dots,\\ 
\g_{x_1 x_1}&=\g_{x_2 x_2}=1-\frac{\rho^{2}}{24}z^{2}+\dots,\\
\g_{x_3 x_3}&=1+\frac{5\rho^{2}}{24}z^{2}+\dots,
\end{split}
\end{align}
If we introduce a temperature, the modification to the metric will start at $O(z^4)$. Further, the scalar field introduces a new scale which breaks scale invariance explicitly and the trace of the boundary stress tensor is now non-zero.
It needs to be checked if the null energy condition is satisfied by the bulk stress tensor $T_{MN}$ given by eq.(\ref{anT}).
Contracting the above with the null vectors $\xi^{\m}$ we have 
\be
T_{MN}\xi^{M}\xi^{N}=\frac{1}{2}[(\partial_{\xi}\phi)^2+e^{2\phi}(\partial_{\xi}\chi)^2],
\ee 
where $\partial_{\xi}(\phi,\chi)=\xi^{M}\partial_{M}(\phi,\chi)$ and $\xi^{M}\xi^{N}g_{MN}=\xi^2=0$. Since the bulk scalar axion follows the profile $\chi=\rho x_3$ then 
\be 
\xi^{M}\partial_{M}\chi=\rho \xi^{x_3},
\ee 
whereas the dilaton field $\phi$ depends on the radial coordinate. The NEC for the bulk stress tensor becomes by contracting with the null vectors $T_{\m\n}\xi^{\m}\xi^{\n}$ as
\begin{align}
\begin{split}
T_{x_3 x_3}=\frac{\rho^2}{2}e^{2\phi}(\xi^{x_3})^2=\frac{\rho^2}{2}e^{2\phi}\geq 0,\\
T_{uu}=\frac{1}{2}(\partial_{\xi}\phi)^2\geq 0\,.
\end{split}
\end{align}
Thus we have explicitly verified that the bulk stress tensor  satisfies the null energy condition.

We now want to verify the calculation for the relative entropy in this low anisotropy regime. As mentioned  before, the low anisotropy phase is thermodynamically unstable. We can thus try to see what happens to the relative entropy in such a phase. Also note that we are considering Einstein gravity for which the  entropy functional is the Ryu-Takayanagi one. Further in the low anisotropy regime, we are interested in, since we are expanding $\gamma_{\m\n}$ upto $O(z^2)$ (assuming a small entangling surface $R\rho\ll 1$) and the stress tensor appears at $O(z^4)$, we have $\Delta H=0$. Here the state $\sigma$ is the vacuum state which corresponds to $\rho=0$ and is conformally invariant. Thus the modular hamiltonian will be the same as in eq.(4). Thus we only need to compute the change in the entanglement entropy. Furthermore, at leading order in $\rho$ we expect to see an inequality and as such we do not need to evaluate the change in the entangling surface.

\noindent Putting in the solution for the entangling surface $ f(x_1,x_2,x_3)=\sqrt{R^2-x_1^2-x_2^2-x_3^3}$ we have 
\be 
\sqrt{h}=\frac{1}{48 (R^2-x_1^2-x_2^2-x_3^3)^{2}R}[48R^2+(R^2-x_1^2-x_2^2-x_3^2)(3R^2-5x_3^2+x_1^2+x_2^2)\rho^2]+O(\rho^4)\,.
\ee 
The entanglement entropy then becomes 
\be S=\frac{2\pi}{\ell_p^3}\int dx_1 dx_2 dx_3  \frac{1}{48 (R^2-x_1^2-x_2^2-x_3^3)^{2}R}[48R^2+(R^2-x_1^2-x_2^2-x_3^2)(3R^2-5x_3^2+x_1^2+x_2^2)\rho^2]\,.
\ee
In spherical polar coordinates $x_3=r \cos\theta, x_1=r\sin\theta \sin\phi,x_2=r\sin\theta \cos\phi$ where $(\theta, \phi)$ are spherical polar coordinates we have 
\be
\Delta_1 S=\frac{2\pi \rho^2}{\ell_p^3}\int \frac{(3R^2-2r^2-3r^2 \cos 2\theta)}{48(R^3-r^2 R)} r^2 \sin\theta d\theta d\phi dr\,.
\ee
Carrying out the $(\theta,\phi,r)$ integrals we find (on reinstating $L$ factors)
\be
\Delta_1 S=\frac{\pi^2 \rho^2 R^2 L^3}{6 \ell_p^3 }(-\frac{5}{3}-\log[\frac{\epsilon}{2R}])\,.
\ee 
Here $\epsilon$ is a cutoff and $r=R-\epsilon$ (since $\epsilon\rightarrow 0$ corresponds to $z\rightarrow 0$ it is related to the UV cutoff). The log-divergence is due to the breaking of conformal invariance by the excited state. However, notice that in the limit of $\epsilon\rightarrow 0$, the result leads to $\Delta_1 S>0$ and hence the positivity of relative entropy is violated. 

Since the positivity of relative entropy in quantum mechanics depends on unitarity (reviewed in appendix A), this leads to the following possible interpretations:
\begin{enumerate}
\item There are additional contributions which we are missing and they are required for the positivity of the relative entropy to hold in this case. One could speculate that there are additional saddle points of the bulk gravity theory which contribute to the entanglement entropy. It will be interesting to find out those saddle points and see if they "unitarize" the problem\footnote{This is very similar to the resolution of information loss paradox in case of eternal $AdS$ Black Holes as formulated by Maldacena \cite{Maldacena:2001kr}. The exponentially small correlation as required by the unitarity arises form the periodically identified Euclidean $AdS$, although this is not the dominant contribution to the canonical ensemble. }. 
\item Holographic relative entropy positivity needs further conditions than just bulk unitarity.  It could be that the derivation of the positivity does not go through in any straightforward manner to quantum field theory. 
\item In the low anisotropy regime, may be there is a loss of bulk unitarity that is not immediately apparent. 
\end{enumerate}
All possibilities need further investigation. Let us first briefly comment on the third possibility.
Expanding the linearized equations near the boundary and upto linear order in $\rho$ we have 
\begin{eqnarray}
(\Box +\frac{2}{L^2})h_{ij}&=&0,\quad 
(\Box+\frac{2}{L^2}) h_{M x_3}+\frac{\rho}{2}L_{M x_3}\chi_1=0,\\
\Box\phi_1-2\rho \partial^{x_3} \chi_1&=&0,\quad
\Box\chi_1=0,\label{scaleq}
\end{eqnarray}
where $h_{MN}$, $\phi_1$ and $\chi_1$ are metric, $\phi$ and $\chi$ fluctuations respectively and $i$, $j$ take values apart from $x_3$. $\nabla_A$ is evaluated using the AdS$_5$ metric. Here $L_{MN}\equiv\delta_{M x_3}\partial_{N}+\delta_{N x_3}\partial_{M}$ is a linear operator. The coupling between the metric and $\chi$ fluctuation is of the form $H h+ L\chi=0, H \chi=0$. But this form is similar to what arises in the context of logarithmic conformal field theories which are non-unitary \cite{grulog}. Thus one should check if there are log modes in the fluctuations. We can do this following \cite{logtown}. According to the arguments in \cite{logtown} log modes arise if the form of the equations is $(\Box+a)^2 h_{\mu\nu}=0$. Let us check what the form of the equations are when we decouple them. Using \footnote{Useful identities can be found for eg. in the appendices of \cite{sss}} $\Box \nabla_A \chi_1=\nabla_A \Box \chi_1-\frac{4}{L^2}\nabla_A \chi_1$, we find that the decoupled equation for $h_{M x_3}$ takes the form 
$$
(\Box+\frac{2}{L^2})(\Box+\frac{4}{L^2})h_{M x_3}=0\,,
$$
while for $\phi$ we get 
$$
\Box(\Box+\frac{4}{L^2})\phi=0\,.
$$
Neither of the four derivative linear operator is of the form $(\Box+a)^2$ and hence following the arguments in \cite{logtown} there are no log modes so the dual field theory is not a log CFT. Naively it may appear that the propagator for say the $\phi$ field will look like $1/(p^2(p^2+m^2))=1/m^2(1/p^2-1/(p^2+m^2))$, and hence the theory is non-unitary. However, this is not true since in addition to the decoupled form of the equations above, the relations in eq.(\ref{scaleq}) still have to hold--any loss of unitarity would have shown up in the asymptotic fall offs in the field. Thus it appears that the other two possibilities become plausible.

In closing this section, we note that in \cite{Relative} it was argued that the relative entropy should increase as the radius of the entangling surface increases. In our case since $\partial_R S(\rho_1|\rho_0)=-\partial_R \Delta_1 S \approx \frac{\pi^2 \rho^2 R L^3}{3 \ell_p^3 }(\log[\frac{\epsilon}{2R}])<0$ and hence this monotonicity would also appear violated.

\section {Smoothness of entangling surface} 

In this section we will derive constraints on the GB coupling by demanding that the entangling surface for sphere, cylinder and the slab close off smoothly in the bulk. The slab case was considered before in \cite{Ogawa:2011fw}. At the onset note that treating the truncated GB gravity on its own leads to problems with entanglement entropy as was pointed out in \cite{Ogawa:2011fw}. In particular if we consider an entangling surface that topologically looks like $\mathcal{M}_2\times R$, then the $\R$ term in the JM entropy functional becomes topological. Adding more handles to the entangling surface will allow us to lower the entanglement entropy arbitrarily if $\lambda>0$. Since this particular sign of $\lambda$ happens to arise in many consistent examples in string theory (see eg.\cite{beyond}), this hints at a problem in interpreting GB gravity on its own as a model for theories describing $c\neq a$--of course, there is no reason to suspect any inconsistencies if this is just the first perturbative correction in an infinite set of higher derivative corrections. We will not have anything to add to this observation. We will simply focus on what constraints arise on the GB coupling by demanding smoothness and compare the result with the causality/positive energy constraints in eq.(\ref{lamcon}).

The general strategy we will adopt is the following. The entangling surface equation follows from eq.(\ref{condition1}). Let us assume that the surface $f(z)$ closes off at $z=z_h$. Around this point, let us assume
\be\label{trial}
f(z)=\sum_{i=0}^\infty c_i (z_h-z)^{\alpha+i}\,.
\ee
We need to determine $\alpha$ and $c_i$'s. At $z=z_h$, $f'(z) \rightarrow +\infty$ since the tangent to the surface will be perpendicular at that point. This means that $0<\alpha<1$ and $c_0>0$. Using these two conditions, we will find that $\lambda$ will be bounded.

\subsubsection*{Cylinder}

Consider the cylinder case first. In cylindrical coordinates, assume the required hypersurface to have the form $r=f(z)$.  From eq.(\ref{condition1}), we get the following equation,
\begin{align}\begin{split}
&\big[z f''(z) \left(6 f_{\infty} \lambda  z f'(z)+f(z) \left((4 f_{\infty} \lambda +1) f'(z)^2-2 f_{\infty} \lambda +1\right)\right)-\left(f'(z)^2+1\right)\\ &\big(f'(z) \left(z (4 f_{\infty} \lambda +1) f'(z)+3 f(z) \left(f'(z)^2-2 f_{\infty} \lambda +1\right)\right)-2 f_{\infty} \lambda  z+z\big)\big]=0\,.\end{split}
\end{align}
We take the trial solution eq.(\ref{trial}) and determine an appropriate $\alpha$.  We obtain $\alpha=1/2, 3/2$. We will drop the second solution since this will lead to a conical tip.
 Expanding the eom in powers of $(z_h-z)$ and setting the leading order term to 0, we get 4 roots of $c_0$. We take the two positive ones,
\be
\sqrt{\frac{2}{3}}\sqrt{z_h(1+4f_{\infty}\lambda\pm\sqrt{1-10f_{\infty}\lambda+16f_{\infty}^2\lambda^2})}\,.
\ee
With $f_{\infty} =(1-\sqrt{1-4\lambda})/2\lambda$, this puts some consraints on $\lambda$. Since the bottom sign vanishes in the $\lambda\rightarrow 0$ limit, we will ignore this solution. For the other case, we have
\be
\hspace{1.5 cm} \lambda\le\frac{7}{64}\,.\ee
The quantities inside the square root have to be positive to make the root real. If we look carefully we will find that  $1-10f_{\infty}\lambda+16f_{\infty}^2\lambda^2$ has to be positive. This is  almost same as that of the tensor channel constraint except  for the extra additional factor of $ 16f_{\infty}^2\lambda^2\,.$ That is why  we get a bigger bound instead of $\lambda \textless \, \frac{9}{100}\,.$

\subsubsection*{Sphere}
The eom reads,
\begin{align}
\begin{split}
&\big[z f''(z) \left(12 f_{\infty} \lambda  z f(z) f'(z)+f(z)^2 \left((4 f_{\infty} \lambda +1) f'(z)^2-2 f_{\infty} \lambda +1\right)+6f_{\infty} \lambda  z^2\right)-\left(f'(z)^2+1\right) \\&\left(6 f_{\infty} \lambda  z^2 f'(z)+2 z f(z) \left((4 f_{\infty} \lambda +1) f'(z)^2-2 f_{\infty} \lambda +1\right)+3 f(z)^2 f'(z) \left(f'(z)^2-2 f_{\infty} \lambda +1\right)\right)\big]=0\,.
\end{split}
\end{align}
We get only $\alpha=1/2$ as a solution to the indicial equation. We get six roots of $c_0$  from the leading order of eom. Three of them are positive:
\be
\sqrt{2z_h}\,,\hspace{0.5 cm}\sqrt{4f_{\infty}z_h\lambda\pm 2\sqrt{2}z_h\sqrt{f_{\infty}\lambda(-1+2f_{\infty} \lambda)}}\,.
\ee
The positivity of the first root cannot give us any constraint on $\lambda$. The other two roots go to zero as $\lambda$ goes to zero so we will ignore them.

\subsubsection*{Slab}
The eom reads,
\be
-3(1-2f_{\infty}\lambda+f'(z)^2)(f'(z)+f'(z)^3)+z(1-2f_{\infty}\lambda+(1+4f_{\infty}\lambda)f'(z)^2)f''(z)=0
\ee
We get  $\alpha=1/2, 1$ which give non-zero $c_0$. Arguing as before we will only consider $\alpha=1/2$.
Here we get the following positive solution for $c_0$:
\be
c_0=\sqrt{\frac{2}{3}}\sqrt{z_h+4f_{\infty}z_h\lambda}\,.
\ee
Demanding this to be positive, we get
\be
-\frac{5}{16}\le \lambda\le \frac{1}{4}\,.
\ee
This agrees with \cite{Ogawa:2011fw}. Thus together with the constraints from the cylinder we have
\be
-\frac{5}{16}\le \lambda\le \frac{7}{64}\,.
\ee
We can recast this inequality as one for $a/c$ where $a, c$ are the Euler and Weyl anomaly coefficients respectively for a 4d CFT. This gives us
\be
\frac{1}{3}\leq \frac{a}{c}\leq \frac{5}{3}\,.
\ee
Quite curiously, the lower bound $1/3$ is precisely what appears in non-supersymmetric theories \cite{hofman, zhibo}, in particular for a free boson. The upper bound of $5/3$ corresponds to a free theory with one boson and two vector fields. For a non-supersymmetric theory, the bound on $a/c$ worked out\footnote{Note $31/18\approx 1.72$ while $5/3\approx 1.67$.} in \cite{hofman} was $1/3\leq a/c \leq 31/18$. Just to point out in words, the $1/3$ came from the cylinder calculation while the $5/3$ came from the slab. The causality constraints on the other hand translates into $1/2\leq a/c \leq 3/2$.

In \cite{ab,abs, area} a different surface equation was proposed for GB gravity which differs from the above considerations at $O(\K^3)$ order. The constraints arising from this are analysed in the appendix. These constraints are weaker than what we found above.  
We compare the different bounds on $\lambda$ in fig.3. As is clear, the causality constraints are the tightest. Note however that the lower bound derived using the method in the appendix coincides with the vector channel constraint. 
 \begin{figure}[htpb]
\centering\includegraphics[scale=.5 ]{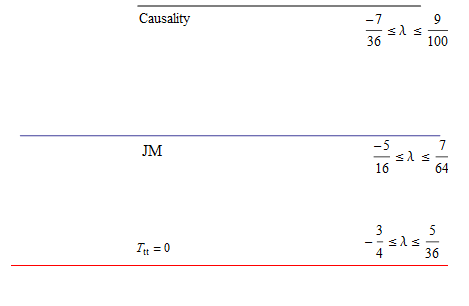}
\caption{Comparison between the various constraints on the GB coupling. The length of the line represents the range of allowed $\lambda$.}

\end{figure}

\section{Discussion}
In this paper we used holographic entanglement to constrain gravity in interesting ways. 
First, we started with the Ryu-Takayanagi entropy functional (which holds for Einstein gravity) and considered what constraints arise at nonlinear order on the metric by demanding that relative entropy is positive. At linearlized level, it is now known that for the spherical entangling surface $\Delta H=\Delta S$ leads to linearized equations for any higher derivative theory of gravity \cite{einstein3}. We considered a constant field theory stress tensor. At the next order, we found interesting constraints on the terms allowed by the positivity of relative entropy. These were more general than what arises from Einstein gravity. We analysed energy conditions for matter that could support these additional theories. We showed that the additional theories could be supported by matter that violates the null energy condition. In other words, holographic relative entropy can be positive although the bulk null energy condition is violated. It is an important open problem to understand if this feature persists for a more general stress tensor. We also gave an example of a model which corresponds to an anisotropic plasma, where for small anisotropy, the relative entropy is negative. This occurred even though the bulk stress tensor satisfied the null energy condition. We gave some possible explanations for this. We will leave further investigations of similar models as an open problem.

Second, we analysed the inequality in Gauss-Bonnet gravity for a given class of small perturbations around the vacuum state. We found that for all our examples, the positivity of the stress tensor two point function ascertained that this inequality was respected. On the bulk side this corresponds to metric fluctuations having positive energy. The simplicity of the final result does cry out for a simpler explanation for our findings. Although the intermediate integrals involved appeared very complicated, the final result was simply proportional to the Weyl anomaly $c$. It would be nice to find a simple explanation for this finding.
It will be interesting to extend our analysis to other higher derivative theories like the quasi-topological gravity \cite{Myers} where we expect the entropy functional to be simple.  Some preliminary studies of the general four derivative theory has been made in appendix C. Another interesting open problem is to consider a disturbance close to the entangling surface. We were able to consider a disturbance that was localized far from the entangling surface and show that the relative entropy is positive. Whether the constraints change as one moves the disturbance closer to the entangling surface is an open problem.

Finally, we also considered other entangling surfaces and demanded that these close off smoothly in the bulk.  In Gauss-Bonnet gravity, this led to the coupling being constrained. The spherical entangling surface did not lead to any constraints on the coupling while the cylindrical and slab entangling surfaces did. This leads to an interesting question. Suppose we knew how to extend the relative entropy results for the spherical entangling surface to other surfaces. Then the smoothness criteria above seems to constrain the coupling of the higher derivative interaction. This suggests that implicitly the relative entropy inequality knows about this. Since the positivity of relative entropy seems to rely only on the unitarity of the field theory, this raises the question if there is any conflict with unitarity if one is outside the allowed region for the coupling. It will be interesting to investigate this question since apriori there does not appear to be any such conflict in the dual gravity. It will also be interesting to find if there are other entangling surfaces which lead to a tighter bound and if the bounds are stronger than the causality constraints.


\vskip 1cm
{\bf Acknowledgments} : We thank Johanna Erdmenger, Diego Hofman, Janet Hung, Rob Myers and Tadashi Takayanagi  for discussions. AS acknowledges support from a Ramanujan fellowship, Govt. of India. AS thanks the organizers of the Annual Taiwan String Workshop and the National Strings Meeting in India for stimulating programs where part of this work was presented.

\appendix
\section{Positivity of relative entropy}
Here we review the proof in quantum mechanics leading to the positivity of relative entropy. This can be found in Nielsen and Chuang's book listed in \cite{nc}. We define relative entropy as,
\be
S(\rho|\sigma)=\hbox{Tr}(\rho \ln \rho)-\hbox{Tr}(\rho\ln \sigma)\,,
\ee
where $\r$ and $\s$ are the density matrices of two different states. Now consider their orthonormal decomposition,
\be
\rho=\sum_i p_i  \ket{i}\bra{i}\hspace{1cm}\mbox{and}\hspace{1cm}\sigma=\sum_j q_j \ket{j}\bra{j}
\ee
where $\ket{i}$ and $\ket{j}$ may not be the same set of eigenvectors. We can write,
\begin{multline}
\hspace{2.8cm}S(\rho|\sigma)=\hbox{Tr}(\rho \ln \rho)-\hbox{Tr}(\rho\ln \sigma)=\sum_i \bra{i}\rho \ln \rho \ket{i}-\sum_i \bra{i}\rho \ln \sigma \ket{i}\\=\sum_i \bra{i}\rho \ln \rho \ket{i}-\sum_i \sum_j \bra{i}\rho \ln \sigma\ket{j}\braket{j|i}=\sum_i p_i \ln p_i - \sum_{i,j} p_i \bra{i}\ln \sigma \ket{j}\braket{j|i}\\=\sum_i p_i \ln p_i - \sum_{i,j} p_i \ln q_j\braket{i|j}\braket{j|i}\ =\ \sum_i p_i \ln p_i - \sum_{i,j} P_{ij}\  p_i \ln q_j\,.
\end{multline}
In the second line we just inserted $1=\sum_j \ket{j}\bra{j}$, and in the last line we have used the notation $P_{ij}=\braket{i|j}\braket{j|i}$. Note that we must have,
\be\label{sum}
\sum_i P_{ij}=\sum_j P_{ij}=1\,.
\ee
Till here, all that we have used is the unitarity of the theory. Now, $\ln x$ is a concave function; which means we must have,
\be
\ln(tx+(1-t)y)\ge t\ln(x)+(1-t)\ln(y)\hspace{1cm}\mbox{for}\hspace{1cm}0\le t \le 1 \,.
\ee
It is easy to generalize this to,
\be
\ln\left(x_1t_1+x_2t_2+...+x_mt_m\right)\ge  t_1\ln(x_1)+t_2\ln(x_2)+...+t_m\ln(x_m)
\ee
$$\mbox{where}\hspace{1cm}\sum_{i=1}^m t_i=1\hspace{1cm}\mbox{and}\hspace{1cm}0\le t_i\le 1\hspace{0.5cm}\forall i\in [1,m]\,.$$
The equality follows if for some $p$, we have $t_p=1$. Using this, and (\ref{sum}) we can write,
\be
-\sum_{j}P_{ij}\ p_i\ln q_j\ge -p_i\ln r_i \hspace{1cm}\mbox{where}\hspace{1cm}r_i=\sum_j P_{ij}q_j\,.
\ee
Hence we get,
\be
S(\rho|\sigma)\ge \sum_i p_i \ln \left( \frac{p_i}{r_i}\right)=- \sum_i p_i \ln \left( \frac{r_i}{p_i}\right)\,.
\ee
Now note that, $\ln x\ge x-1$. This gives 
$$
S(\rho|\sigma)\ge - \sum_i p_i \ln \left( \frac{r_i}{p_i}\right)\ge- \sum_i p_i \left(1- \frac{r_i}{p_i}\right)\,,
$$
\be
=\sum_i (p_i-r_i)=0\,.
\ee
Hence,  $S(\rho|\sigma)\ge 0$ and the equality follows when $\rho=\sigma$. To repeat, the only assumption that went in the proof was the unitarity of the quantum theory. So, whenever we have a unitary theory we can expect relative entropy to be positive.

\section{Smoothness conditions arising from $T_{tt}=0$}

 As shown in \cite{area, ab} setting the time-time component of the Brown-York stress tensor($T_{tt}$) for the Einstein gravity to zero on the co-dimension one surface $r=f(z)$ we can get the shape of the extremal surface.  This can be extended to the Gauss-Bonnet case \cite{ab}. We do the same analysis setting $T_{tt}=0$ to find $f(z)$ as a series expansion around $z=z_{h}\,.$ Then we proceed in the similar way to get the bound on $\lambda$ demanding the positivity of the coefficient of the leading term of $f(z)\,.$  We do this analysis for the three cases mentioned below. 
The general surface equation is given by \cite{ab}
\be \label{condition2}
 \K+\lambda L^{2}\big[(\R\K-2 \R_{ij}\K^{ij})+\frac{1}{3}(-\K^{3}+3 \K \K_{2}-2\K_{3})\big]=0\,.\\
\ee
Here $\K_2=\K_a^b \K_b^a$ and $\K_3=\K_a^b \K_b^c \K_c^a$. The equations from the JM functional and the above proposal differ at $O(\K^3)$, namely the $O(\K^3)$ terms are absent in the former. 

\paragraph {Sphere:}
 Setting $T_{tt}=0$ we get,
\begin{align}
\begin{split}
\big(f'(z)^2+1\big) \big(4 f_{\infty} \lambda  z^2 f'(z)+2 z f(z) \big((2 f_{\infty} \lambda +1) f'(z)^2-2 f_{\infty} \lambda +1\big)+f(z)^2 f'(z) \big((3-2 f_{\infty} \lambda ) f'(z)^2\\ -6 f_{\infty} \lambda +3\big)\big)-z f''(z) \big(8 f_{\infty} \lambda  z f(z) f'(z)+f(z)^2 \big((2 f_{\infty} \lambda +1) f'(z)^2-2 f_{\infty} \lambda +1\big)+4 f_{\infty} \lambda  z^2\big)=0\,.\\
\end{split}
\end{align}
Putting $$f(z)=c_0 (z_h-z)^{\alpha}+c_1(z_h-z)^{\alpha+1}+\cdots$$   and solving order by order we get $\alpha=\frac{1}{2}$ and $c_{0}=\sqrt{2 z_{h}}$ as before.  We neglect the other root as it vanishes in $\lambda=0$ limit. So this will not give any bound on $\lambda\,.$
\paragraph{Cylinder:} In this case the eom reads
\begin{align}
\begin{split}
&\big(f'(z)^2+1\big) \big(f'(z) \big(z (2 f_{\infty} \lambda +1) f'(z)+f(z) \big((3-2 f_{\infty} \lambda ) f'(z)^2-6 f_{\infty} \lambda +3\big)\big)-2 f_{\infty} \lambda  z+z\big)\\ &-z f''(z) \big(4 f_{\infty} \lambda  z f'(z)+f(z) \big((2 f_{\infty} \lambda +1) f'(z)^2-2 f_{\infty} \lambda +1\big)\big)=0\,.
\end{split}
\end{align}
For this case we get two solutions as before. For $\alpha=\frac{1}{2}$ we get $$ c_{0}=\sqrt{2} \sqrt{\frac{z_{h} \left(\sqrt{12 f_{\infty}^2 \lambda ^2-8 f_{\infty} \lambda +1}+1+2 f_{\infty} \lambda \right)}{3-2 f_{\infty} \lambda }}\,.$$ We have only considered the root which is continuously connected to the Einstein case. From $c_{0}>0$ we get 
\be
 \lambda\,\,\leq\,\,  5/36\,,
\ee
 which is same as that of the condition $a\geq  0\,.$ Here $a$ is the Euler anomaly. 
\paragraph{Rectangular strip:} The eom is
\begin{align}
\begin{split}
\left(f'(z)^3+f'(z)\right) \left(-\left((2 f_{\infty} \lambda -3) f'(z)^2+6 f_{\infty} \lambda -3\right)\right)-z f''(z) \left((2 f_{\infty} \lambda +1) f'(z)^2-2 f_{\infty} \lambda +1\right)=0\,.
\end{split}
\end{align}
From this we get $\alpha=\frac{1}{2}$ and $ c_{0}=\frac{\sqrt{z_{h} (2 f_{\infty} \lambda +1)}}{\sqrt{-f_{\infty} \lambda +\frac{3}{2}}}\,.$ Demanding the positivity we finally obtain 
\be
-\frac{3}{4}\le \lambda \leq \frac{1}{4}\,.
\ee
This condition ensures  $c>0$ where $c=1-2f_{\infty}\lambda$ is the Weyl anomaly.
 From this we get the following bound on $\lambda $:
\be
 -\frac{3}{4} \leq \lambda  \leq \frac{5}{36}\,.
\ee
In terms of $a,c$, this translates into $0\leq a/c \leq 2$.
\section{Relative entropy for $R^2$ theory in  shockwave background}
In this section we want to sketch the calculation for the relative entropy in shockwave background for a general $R^2$ theory\footnote{The corresponding entropy functional will be useful in studying relative entropy in non-unitary log CFTs--for recent applications for entanglement entropy in these theories, see \cite{alilog}.} where the disturbance is located very far away from the entangling surface. The action for this theory is shown below,
\be
I=\int d^5x\sqrt{G}\left(R+\frac{12}{L^2}+\frac{L^2}{2}\left(\lambda_{3}R^2+\lambda_{2}R_{AB}R^{AB}+\lambda_{1}R_{ABCD}R^{ABCD}\right)\right)\,.\\
\ee
In this case, $f_\infty$ satisfies $1-f_\infty+\frac{1}{3}f_\infty^2 (\lambda_1+2\lambda_2+10\lambda_3)=0$.
  We start of  with the shockwave metric as given in eq.(\ref{new})\,. We have explicitly checked that this is still a solution for the $ R^{2}$ theory. Next we quote the area functional for this theory \cite{solo,abs,dong},
\be \label{areaR}
S_{EE}=\frac{2\pi}{\lp^{3}}\int d^{3}x \sqrt{h} \big(1+\frac{L^{2}}{{2}}(2\lambda_{3}R+\lambda_{2}( R_{AB}n^{A}_{i}n^{B}_{i}-\frac{1}{2} \K^{i}\K_{i})+2\lambda_{1} (R_{ABCD}n^{A}_{i}n^{B}_{j}n^{C}_{i}n^{D}_{j}-\K^{i}_{ab}\K^{ab}_{i}))\big)\,.\\
\ee 
Here $i$ denotes the two transverse directions to the co-dimension 2 surface $z=f(x_1,x_2,x_3)$ and $ t=0$ and $\K_{i}$'s are the two extrinsic curvatures along these two directions pulled back to the surface and $a,b$ are three dimensional indices. Then we proceed in the same way as before. We  set $z=z_{0}=\sqrt{R^2-r^2}\,.$  Also as before we set $x_1'=0$ and without loss of any generality and we will  expand the integrand around $x_2'=\infty$ .
First we expand upto $O(\epsilon^3)$ which is the linearized term and hence should yield $\Delta H=\Delta S$. The expression for $\Delta^{(1)}S$ is 
\be
\Delta^{(1)}S=\frac{16 \pi^2 L^5 R^4}{15 f_\infty^{5/2}\lp^3 x_2'^6}(1+2 f_\infty(\lambda_1-2(\lambda_2+5\lambda_3))\,.
\ee
The $\lambda_i$ dependence has packaged into being proportional to $c$ for the general theory \cite{higher}. Using the results of \cite{einstein3} (eq.(6.29) in that paper with the typo mentioned in footnote 4 taken into account), we find that $\Delta H=\Delta S$ at this order as expected.
Then we expand (\ref{areaR}) upto $\epsilon^4$  order and pick out the $\epsilon^4$ term which gives us the $\Delta^{(2)} S\,.$  Note that for a general $R^2$ theory the surface term is not known. So we can only do this calculation for the disturbance located very far away from the entangling surface such that we do not have  to consider the perturbation to the entangling surface as this will contribute to some order higher than $\epsilon^{4}$. Further since the extrinsic curvatures are each proportional to $\epsilon^3$ and hence the $O(\K^2)$ terms would be proportional to $O(\epsilon^6)$, they will not contribute. The result before carrying out the integrations is shown below,
\begin{align}
\begin{split}
\Delta^{(2)} S&=\frac{2\pi}{\lp^{3}}\int dx_{3} dx_{1} dx_2\big[\frac{3 L^5}{2 f_{\infty}^{5/2} R x_{2}'^8}( (x_1^2+x_2^2+x_3^2-R^2) [R^2 (2 f_{\infty} (23 \lambda_{1}+\lambda_{2}-10 \lambda_{3})+1)\\&-60 f_{\infty} \lambda_{1} \left(x_1^2+x_2^2\right)-x_3^2 (2 f_{\infty} (23 \lambda_{1}+\lambda_{2}-10 \lambda_{3})+1)])\big]\,.
\end{split}\end{align}
Then we perform the integration over $x_3$ which goes from $-\sqrt{R^{2}-r^2}$ to $\sqrt{R^2-r^2}$ and 
$ x_{1}= r \cos(\theta)\,\,, \,\,x_{2}= r \sin(\theta)\,.$ Now after some algebraic manipulation we can write the integrand as,
\be
\Delta^{(2)}S=-\frac{48 \pi ^2 L^5 R^6 (1+2 f_{\infty} (13 \lambda_{1}+\lambda_{2}-10 \lambda_{3}))}{35 f_{\infty}^{5/2} \lp^3 x_2'^8}\,.\\
\ee
Note that this is not proportional to $c$ for this theory. Since for generic values of the couplings $\lambda_i$, the bulk theory is non-unitary this may not be surprising. This may be indicative of the fact that rather than depending only on the two point function of the stress tensor, the higher point functions also contribute as in the second reference in \cite{hofman}. The bulk theory will make sense as an effective theory where the couplings are small. In this circumstance, we can use field redefinitions to make the theory equivalent to Gauss-Bonnet with $\lambda\propto \lambda_1$. 
For the Gauss-Bonnet value $\lambda_{1}=\lambda_{3}=\lambda\,, \lambda_{3}=-4\lambda$ it reduces to,
\be
\Delta^{(2)}S=-\frac{48 \pi ^2 L^5 R^6 (1-2 f_{\infty} \lambda )}{35 f_{\infty}^{5/2} \lp^3 x_2'^8}\,,
\ee
which is proportional to $c$ for the GB theory.

\singlespacing

\end {document}